\documentclass[aps,pra,floatfix,twocolumn,superscriptaddress]{revtex4-1}
\usepackage{graphicx}
\usepackage[english]{babel}
\usepackage{amsmath}
\usepackage{amssymb}
\usepackage{tensor}

\newcommand{\be}{\begin{eqnarray}}
\newcommand{\ee}{\end{eqnarray}}
\newcommand{\p}{\partial}

\newcommand{\dc}{c^{\dagger}}

\def\ket#1{|#1\rangle}
\def\bra#1{\langle #1 |}
\def\ep#1{\langle #1 \rangle}

\begin{document}

\title{Non-Hermitian generalizations of extended Su-Schrieffer-Heeger models}

\author{Yan He}
\affiliation{College of Physics, Sichuan University, Chengdu, Sichuan 610064, China}
\email{heyan$_$ctp@scu.edu.cn}

\author{Chih-Chun Chien}
\affiliation{Department of physics, University of California, Merced, CA 95343, USA.}
\email{cchien5@ucmerced.edu}

\begin{abstract}
Non-Hermitian generalizations of the Su-Schrieffer-Heeger (SSH) models with higher periods of the hopping coefficients, called the SSH3 and SSH4 models, are analyzed. The conventional construction of the winding number fails for the Hermitian SSH3 model, but the non-Hermitian generalization leads to a topological system due to a point gap on the complex plane. The non-Hermitian SSH3 model thus has a winding number and exhibits the non-Hermitian skin effect. Moreover, the SSH3 model has two types of localized states and a zero-energy state associated with special symmetries. The total Zak phase of the SSH3 model exhibits quantization, and its finite value indicates coexistence of the two types of localized states. Meanwhile, the SSH4 model resembles the SSH model, and its non-Hermitian generalization also exhibits the non-Hermitian skin effect. A careful analysis of the non-Hermitian SSH4 model with different boundary conditions shows the bulk-boundary correspondence is restored with the help of the generalized Brillouin zone or the real-space winding number. The physics of the non-Hermitian SSH3 and SSH4 models may be tested in cold-atom or other simulators.
\end{abstract}

\maketitle

\section{Introduction}
The Su-Schrieffer-Heeger (SSH) model~\cite{SSH79} has been a paradigm of one-dimensional topological insulators~\cite{Asboth2016,Chiu2016}. In the simplest version, the SSH model describes non-interacting quantum particles hopping in a one-dimensional (1D) lattice with alternating hopping coefficients. The bulk-boundary correspondence of the SSH model~\cite{Asboth2016} shows that with periodic boundary condition, the winding number serves as a topological invariant differentiating the two topologically distinct regimes determined by the ratio of the two hopping coefficients. With open boundary condition, localized edge states can be found at the ends of the system. Importantly, the number of edge states can be determined by the winding number. Originally proposed for polyacetylene~\cite{SSH79}, the SSH model has been demonstrated experimentally by cold-atoms in optical superlattices~\cite{ZakPhase} and by chlorine atoms on copper surface~\cite{Drost17} and many other quantum systems. Classical mechanical systems may also mimic the SSH model~\cite{Huber16,ChienPRB18}.

There have been many generalizations of the SSH model. By considering the SSH model as a system with a periodic pattern of the hopping coefficients, there are two sites per unit cell due to the alternating hopping coefficients, so the period is two. One line of generalizations considers the effects of increasing the period of the patterns of hopping, and the models are usually known as the extended SSH models. Here we call the extended SSH models with period-3 and period-4 hopping coefficients the SSH3 and SSH4 models, respectively.  Interestingly, the Hermitian SSH3 model defies the conventional definition of the winding number~\cite{Chiu2016, Asboth2016} for 1D systems with chiral symmetry. Moreover, the zero-energy state of the SSH3 model is associated with a symmetry of odd-number lattices instead of the band-topology~\cite{HePRA18}. In contrast, the SSH4 model~\cite{SSH4} is a topological insulator, and it is basically the SSH model in disguise. The SSH4 model has the chiral symmetry and belongs to the same class of the SSH model, and the winding number can characterize its band topology. The SSH4 model has four bands with more mid-gap states located inside the three gaps. However, only the zero-energy state is protected by the chiral symmetry and associated with the band-topology while the other mid-gap states are associated with a special symmetry. In addition, the SSH4 model has a much larger parameter space and can display richer phenomena. While the SSH4 model has been demonstrated in cold-atom experiments~\cite{Xie19}, similar experiments are expected to realize the SSH3 model as well~\cite{HePRA18}.

On the other hand, the SSH model has a Hermitian Hamiltonian. The formulation of non-Hermitian quantum mechanics~\cite{Bender07,Ganainy18,Ashida20} has introduced another line of generalizations of the SSH model. The non-Hermitian SSH model has been intensely studied~\cite{Wang1,Wang2,Yuce19,Yokomizo19,KunstPRL18,Leykam17,LeePRB19} and, just like the paradigmatic Hermitian SSH model, become an important platform for investigating non-Hermitian topological phenomena. The results from the systems with periodic and open boundary conditions may no longer agree, and the introduction of the biorthonormal basis and generalized Brillouin zone are crucial in restoring the bulk-boundary correspondence in the non-Hermitian SSH model~\cite{Wang1,Wang2,KunstPRL18,Leykam17,Imura19,Koch19,Borgnia20}. Moreover, the presence of asymmetric hopping coefficients between the same pair of sites leads to the non-Hermitian skin effect, where the bulk states exhibit skewed profiles~\cite{TorrePRB18,Wang1,SongPRL19}.
The classifications of non-Hermitian topological systems are also different from those of Hermitian topological systems~\cite{Li19,Wojcik19,Xi19,Bessho19,Zhou19,Ghatak19,YoshidaPRB19,YoshidaNHMech}.

Here we integrate the two lines of generalizations of the SSH model and investigate non-Hermitian SSH3 and SSH4 models. While the Hermitian SSH3 model is topologically trivial, a generalization to the non-Hermitian model leads to topological properties as the eigenstates encircles the origin on the complex plane, illustrating how non-Hermitian generalizations can change the physics of the Hermitian counterpart. Moreover, we will show two types of localized states not associated with the topology as well as a zero-energy state from a symmetry of the non-Hermitian SSH3 model. On the other hand, the SSH4 model is already topological in the Hermitian case. Nevertheless, a non-Hermitian generalization of the SSH4 model shows the non-Hermitian skin effect, causing the skewed profiles of the bulk states. In addition, the energy spectrum and topological invariant of the non-Hermitian SSH4 model become sensitive to the boundary conditions, showing the typical behavior of non-Hermitian systems~\cite{Wang1,Wang2}. There has been previous work on a non-Hermitian generalization of the SSH3 model~\cite{JinPRA17}, but the system has diagonal non-Hermitian terms rather than the off-diagonal non-Hermitian terms of the systems discussed here, leading to different physics.

The non-Hermitian SSH3 model will be shown to have a point gap, as all of its eigenvalues on the complex plane encircle a given point. Non-Hermitian models with point gaps have been studied in Ref.~\cite{Gong18}. An important feature of the models with point gaps is that the 10-fold way classification of the Hermitian models collapses into a 6-fold way classification of the non-Hermitian models. While the A class of Hermitian systems in 1D is topologically trivial, one can find a $\mathcal{Z}$ topological index for the class A in 1D because of the change of the classification if the non-Hermitian model has a point gap.  In such a case, the full spectrum is usually very sensitive to the boundary condition, giving rise to certain exotic bulk-boundary correspondence without a Hermitian counterpart. On the other hand, the non-Hermitian SSH4 model will be shown to have a line gap because its eigenvalues form several clusters that  can be separated by lines on the complex plane. For models with line gaps, the usual 10-fold way classification of the Hermitian models is refined to a much more complicated classification \cite{Kawabata}. As for the SSH4 model, the non-Hermitian generalization studied here resembles the Hermitian one. Nevertheless, the non-Hermitian skin effects will lead to quantitatively different results with different boundary conditions, so additional analyses are performed to restore the bulk-boundary correspondence of the non-Hermitian SSH4 model.

The rest of the paper is organized as follows. Section~\ref{sec:SSH3} summarizes the Hermitian SSH3 model and then presents its non-Hermitian generalization. We discuss the topological properties characterized by the energy spectrum and winding number and analyze two localized edge states and a zero-energy state not associated with the topology. The Zak phase will also be analyzed. Section~\ref{sec:SSH4} reviews the Hermitian SSH4 model and then presents its non-Hermitian generalization. The results from the system with different boundary conditions differ from each other due to the non-Hermitian skin effect. We use the generalized Brillouin zone and real-space winding number to show that the bulk-boundary correspondence works for the non-Hermitian SSH4 model. Possible realizations of the non-Hermitian SSH3 and SSH4 models in ultracold atoms or other types of simulators are discussed in Section.~\ref{sec:exp}. Finally, Section~\ref{sec:conclusion} concludes our work.

\section{SSH3 model}\label{sec:SSH3}
\subsection{Hermitian model}
We consider the Hermitian model with period-3 hopping coefficients, or the SSH3 model. This can be thought of as a generalization of the SSH model with three lattice sites in one unit cell. The real-space Hamiltonian is
\be
H=\sum_{j}\Big[t_1 \dc_{j,1}c_{j,2}+t_2 \dc_{j,2}c_{j,3}+t_3 \dc_{j,3}c_{j+1,1} + h.c.\Big].
\ee
Here the hopping coefficients $t_{1,2,3}$ are assumed to be real.
In momentum space, the Bloch Hamiltonian is a $3$ by $3$ matrix:
\be
H=\left(
    \begin{array}{ccc}
      0 & t_1 & t_3 e^{-ik_x} \\
      t_1 & 0 & t_2 \\
      t_3 e^{ik_x} & t_2 & 0
    \end{array}
  \right).\label{P3}
\ee
Here $k_x$ is the crystal momentum.

In general, a 1D tight-binding model with nearest neighbor hopping has chiral (sublattice) symmetry. One can see this by considering a generic model given by
$H_G=\sum_j (t_j\dc_{j} c_{j+1}+t_j\dc_{j+1}c_j)$
with some arbitrary hopping coefficients $t_j$ for $j=1,\cdots,N$. If one transforms $c_j\to (-1)^jc_j$, then $H_G\to-H_G$, which manifests the chiral symmetry. For a model with periodic boundary condition, chiral symmetry also requires the total number of sites to be even. Nevertheless, the chiral symmetry of the SSH3 model is obscured in momentum space, as shown in Eq.~(\ref{P3}). It is possible to make the chiral symmetry more transparent by putting two adjacent unit-cells of the SSH3 model together to construct a 6-band model as
\be
H=\left(
    \begin{array}{cccccc}
    0 & t_1 & 0& 0 & 0 & t_3 e^{-ik_x} \\
    t_1 & 0 & t_2 & 0 &0 & 0\\
    0 & t_2 & 0 & t_3 & 0 &0\\
    0 & 0& t_3 & 0 & t_1 & 0\\
    0 & 0 & 0 & t_1 & 0 & t_2\\
    t_3 e^{ik_x} & 0 & 0 & 0 & t_2 & 0
    \end{array}
  \right).\label{P6}
\ee
This way, the chiral symmetry manifests itself as $\Gamma H\Gamma=-H$ with $\Gamma_{ij}=(-1)^i\delta_{ij}$.

Eq.~(\ref{P6}) can be cast into an off-diagonal form as
\be
SHS^{-1}=\left(
    \begin{array}{cc}
      0 & g \\
      g^{\dag} & 0
    \end{array}
  \right),\,
g=\left(
    \begin{array}{ccc}
      t_1 & 0 & t_3 e^{-ik_x} \\
      t_2 & t_3 & 0 \\
      0 & t_1 & t_2
    \end{array}
  \right).
\ee
The conventional way of characterizing the topology of 1D systems with chiral symmetry is to count how many times the complex number det($g$) winds around the origin of the complex plane as $k$ moves from $0$ to $2\pi$ \cite{Chiu2016}.  However, it can be show that det$(g)=t_1t_2t_3(1+e^{ik})$ for the SSH3 model. As a consequence, det($g$) passes through the origin on the complex plane as $k$ moves from $0$ to $2\pi$ for any choice of the parameters. Thus, the winding number is not well defined for the SSH3 model. Importantly, the crossing of the origin of det($g$) has nothing to do with the gap closing points of the SSH3 model. This is in stark contrast to the ordinary SSH model, where det$(g)=0$ only occurs at the gap closing point.  Therefore, the chiral symmetry does not lead directly to nontrivial band topology of the SSH3 model. We mention that an extension of Eq.~(\ref{P3}) to a 2D model can result in nonzero Chern numbers by introducing a fictitious periodic momentum $k_y$ to modulate the hopping coefficients~\cite{HePRA18}.

\begin{figure}
\centerline{\includegraphics[width=0.6\columnwidth]{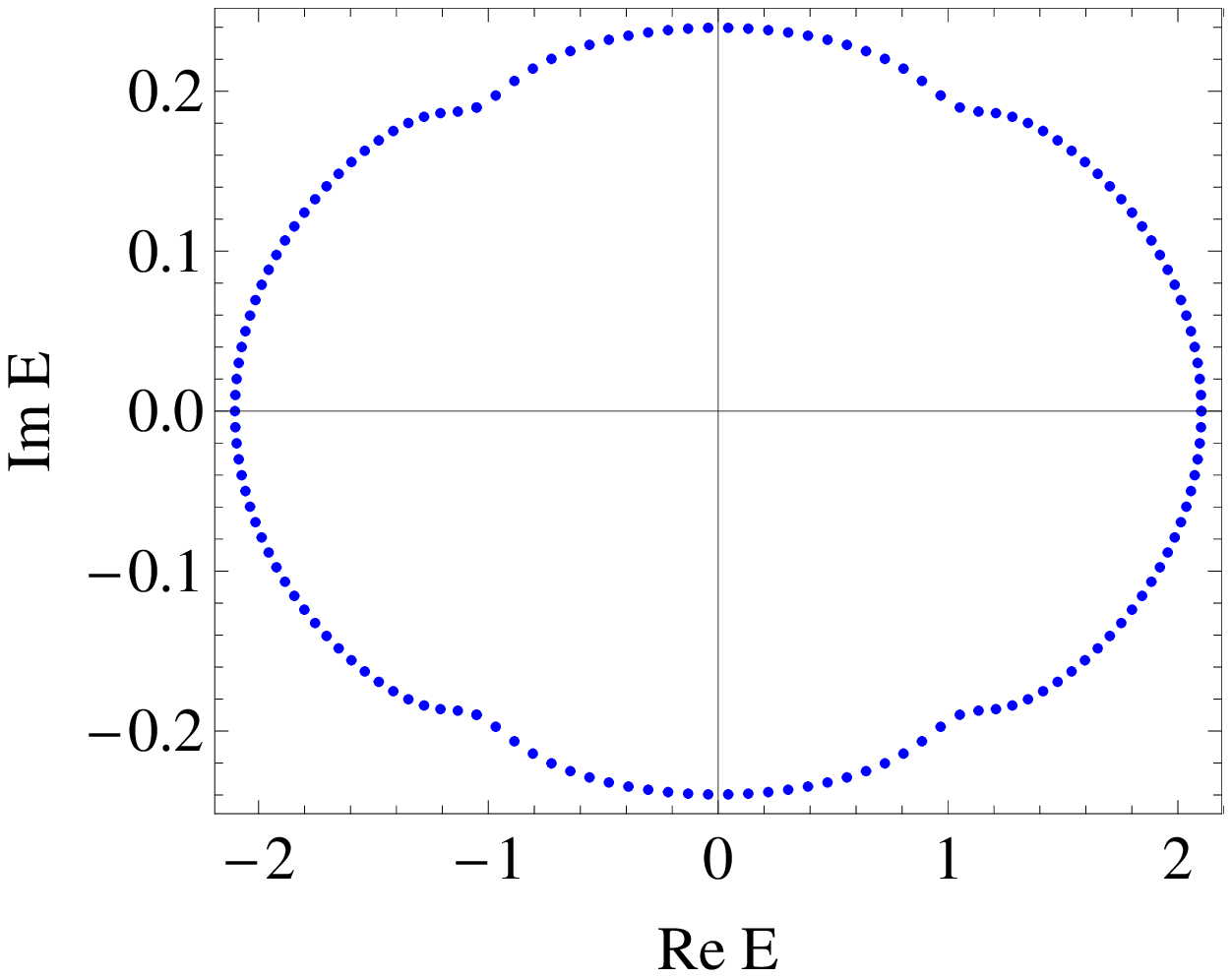}}
\caption{Energy spectrum of the non-Hermitian SSH3 model~(\ref{P3-nh}) on the complex plane. Here $t_1=1.2$, $t_2=t_3=1$, $\gamma=0.4$, and periodic boundary condition has been used.}
\label{E-C}
\end{figure}

Nevertheless, the chiral symmetry protects a zero-energy state of the SSH3 model with open boundary condition when $N$ is odd. This actually applies to any tight-binding model with only nearest-neighbor hopping. By construction the matrix $V=diag\{1,-1,1,-1,\cdots,-1,1\}$, one finds $VHV=-H$. If $\psi$ is an eigenvector of $H$ with the eigenvalue $E$, then $V\psi$ is an eigenvector of $H$ with the eigenvalue $-E$. When the dimension of $H$ is odd, there must be a zero-energy state. In the context of fermions in odd-number lattices, the zero-energy state has been discussed in Ref.~\cite{Semenoff20} for Hermitian models. The zero-energy state may contribute a peak to the local density of state~\cite{HePRA18}. We remark the the zero-energy state is not necessarily a localized state. However, in the presence of the non-Hermitian skin effect (discussed later), the profile of the zero-energy state of the non-Hermitian SSH3 model skews towards one end. 

\subsubsection{Symmetry-protected localized states}
The Hermitian SSH3 model with open boundary condition supports two types of localized states, which are protected by a symmetry. For a finite-size chain, the Hamiltonian is
\be
H=\left(
    \begin{array}{cccccc}
      0 & t_1 & & & & \\
      t_1 & 0 & t_2 & & & \\
       & t_2 & 0 & t_3 &  & \\
       &  & t_3 & \ddots & \ddots &  \\
       &  &  & \ddots & 0 &t_2\\
       &  &  &  & t_2 &0
    \end{array}
  \right).\label{H-r}
\ee
From the eigenvalue equation $H\psi=E\psi$ with $\psi=(a_1,\dots,a_N)^T$, we find
\be
&&t_1 a_{3n-2}-E a_{3n-1}+t_2 a_{3n}=0,\\
&&t_2 a_{3n-1}-E a_{3n}+t_3 a_{3n+1}=0,\\
&&t_3 a_{3n}-E a_{3n+1}+t_1 a_{3n+2}=0.
\ee
Here $n=1,\cdots,M$ and $M=N/3$. The open boundary condition leads to
\be
&&-Ea_1+t_1 a_2=0,\\
&&t_2a_{3M-1}-E a_{3M}=0.
\ee

The two types of localized states correspond to two sets of solutions to the eigenvalue problem. We first look for a solution with $E=t_2$, which has the following form
\be
a_{3n-1}=a_{3n}=\Big(-\frac{t_1}{t_3}\Big)^{M-n},\quad a_{3n-2}=0.
\ee
If $t_1<t_3$, then the amplitude $a_{3n-1}=a_{3n}$ exponentially decreases as $n$ decreases. Under this condition, the boundary condition $a_2=0$ is satisfied in the thermodynamic limit and is approximately satisfied in a finite chain. Therefore, we have found the eigenstate with $E=t_2$. Similar calculations lead to the eigenstate with $E=-t_2$. The profiles of both states are localized at the right end of a finite chain. We will call the $E=\pm t_2$ states the type-I localized states.

Next, we look for the eigenstate with $E=t_1$. Its amplitude satisfies
\be
a_{3n-2}=a_{3n-1}=\Big(-\frac{t_2}{t_3}\Big)^{n-1},\quad a_{3n}=0.
\ee
If $t_2<t_3$, then $a_{3n-2}=a_{3n-1}$ exponentially decrease as $n$ increases. The boundary condition $a_{3M-1}=0$ is satisfied in the thermodynamic limit and is approximately satisfied in a finite chain. Thus, the eigenstate with $E=t_1$ has been found. Similar calculations lead to the eigenstate with $E=-t_1$. Their profiles are localized at the left end of a finite chain. The $E=\pm t_1$ states will be called the type-II localized states. We mention that when the number of sites of a finite chain satisfies $N=3m+2$, the type-II localized states becomes an exact solution and has been discussed in Ref.~\cite{HePRA18}.

The two sets of localized states are associated with a special symmetry of the Hamiltonian (\ref{H-r}). Since the Hermitian SSH3 model does not have a well-defined topological invariant, those localized states are not associated with the band-topology.
We first analyze the type-I localized states at $E=\pm t_2$ by introducing a diagonal matrix $\Gamma'=\mbox{diag}\{-1,1,1,-1,1,1,\cdots,1\}$. Then it follows that
\be
&&\Gamma'H\Gamma'+H=2H_0\label{H0},\\
&&H_0=\mbox{diag}\{0,t_2\sigma_2,\cdots,0,t_2\sigma_2\}.\nonumber
\ee
To see the symmetry leads to the state, we consider a wave function $\psi=(a_1,\cdots,a_N)^T$ satisfying $a_{3j-2}=0$ and $a_{3j-2}=a_{3j}$. It can be verified that $\Gamma'\psi=\psi$ and $H_0\psi=t_2\psi$. If one assumes that $\psi$ is an eigenvector of $H$ and makes use of Eq.~(\ref{H0}), the eigenvalue must be $E=t_2$. A similar analysis shows the other type-I localized state $E=-t_2$ follows the same kind of symmetry. Moreover, the type-II localized state also follows a similar analysis by introducing $\Gamma''=\mbox{diag}\{1,1,-1,\cdots,-1\}$. Then we have
\be
&&\Gamma''H\Gamma''+H=2H_0,\\
&&H_0=\mbox{diag}\{t_1\sigma_1,0,\cdots,t_1\sigma_1,0\}.\nonumber
\ee
Following similar arguments, one can see the type-II localized states are associated with the symmetry.

\subsection{Non-Hermtian generalization and winding number}
Next, we introduce a non-Hermitian generalization of the SSH3 model. With periodic boundary condition, the 1D Bloch Hamiltonian is given by
\be
H=\left(
    \begin{array}{ccc}
      0 & t_1+\gamma & t_3 e^{-ik} \\
      t_1-\gamma & 0 & t_2 \\
      t_3 e^{ik} & t_2 & 0
    \end{array}
  \right).\label{P3-nh}
\ee
Here $\gamma$ is a real-valued parameter. We remark that the topology of the non-Hermitian model~\eqref{P3-nh} is characterized by both the eigenvectors and eigenvalues. Since the eigenvalues of a non-Hermitian Hamiltonian can be complex-valued, interesting topology may arise when the eigenvalues of a model encircle a point on the complex plane. To characterize the topological properties, the following definition of the winding number has been introduced~\cite{Gong18}.
\be
W=\frac{1}{2\pi i}\int_0^{2\pi}dk \frac{d\ln(\det H)}{dk}.
\ee

Figure \ref{E-C} shows the eigenvalues of Eq.~(\ref{P3-nh}) on the complex plane. We assume $t_1=1.2$, $t_2=t_3=1$ and $\gamma=0.4$. One can see that the eigenvalues surround the origin of the complex plane, exhibiting an energy spectrum with a point gap. From Eq.~(\ref{P3-nh}), we have
\be
\mbox{det}(H)=2t_2t_3(t_1\cos k+i\gamma\sin k)\equiv r(k)e^{i\phi(k)}.
\ee
In contrast, $\det(H)$ is real-valued for the Hermitian SSH3 model, so there is no winding. For the non-Hermitian SSH3 model, 
\be
W&=&\frac{1}{2\pi i}\int_0^{2\pi}dk\Big(\frac{d}{dk}\ln r(k)+e^{-i\phi}\frac{d}{dk}e^{i\phi}\Big)\nonumber\\
&=&\mbox{sign}(\gamma).
\label{wind}
\ee
Characterized by the winding number, the 1D non-Hermitian SSH3 model is topologically non-trivial. It can be shown that as long as all of the parameters $t_{1,2,3}$ and $\gamma$ are nonzero, $\det(H)$ is always a non-zero complex number. As $k$ moves from $0$ to $2\pi$, $\det(H)$ will circle around the origin of the complex plane, giving rise a non-zero wind number. Therefore, the introduction of the non-Hermitian term changes the topologically trivial Hermitian SSH3 model to a topological but non-Hermitian one.
Interestingly, similar behavior can be observed even when $t_1=t_2=t_3=1$ in the non-Hermitian model with nonzero $\gamma$. 

We would like to mention that this non-Hermitian SSH3 model is very similar to the Hatano-Nelson model \cite{Hatano}, which is a single band model with asymmetric hopping coefficients. In both models, there is a point gap in the complex energy spectrum, but the Hermitian counterparts are topologically trivial. Since the winding number has no Hermitian counterpart, those non-Hermitian models with point gaps may exhibit unconventional behavior. For example, the spectrum and eigenvectors may experience a sudden change when one changes the boundary condition from periodic to open, or some exponentially small change of the boundary condition may lead to an order-one change of the spectrum \cite{Xiong}. 

We remark that the non-Hermitian term of the SSH3 model~\eqref{P3-nh} is only on one of the three sites in a unit cell, so the non-Hermitian SSH3 model with equal hopping coefficients still differs from the Hatano-Nelson model. Moreover, in semi-infinite non-Hermitian systems with point gaps, there may be localized states associated with the winding number~\cite{OkumaPRL20}. Since we are considering finite-size systems that may be studied in quantum simulators, there is no localized states associated with the winding number of the non-Hermitian SSH3 model. On the other hand, Ref.~\cite{OkumaPRL20} shows the winding number is the origin of the non-Hermitian skin effect in systems with point gaps. The non-Hermitian skin effect of the non-Hermitian SSH3 model will be presented shortly.

\begin{figure*}
\centerline{\includegraphics[width=\textwidth]{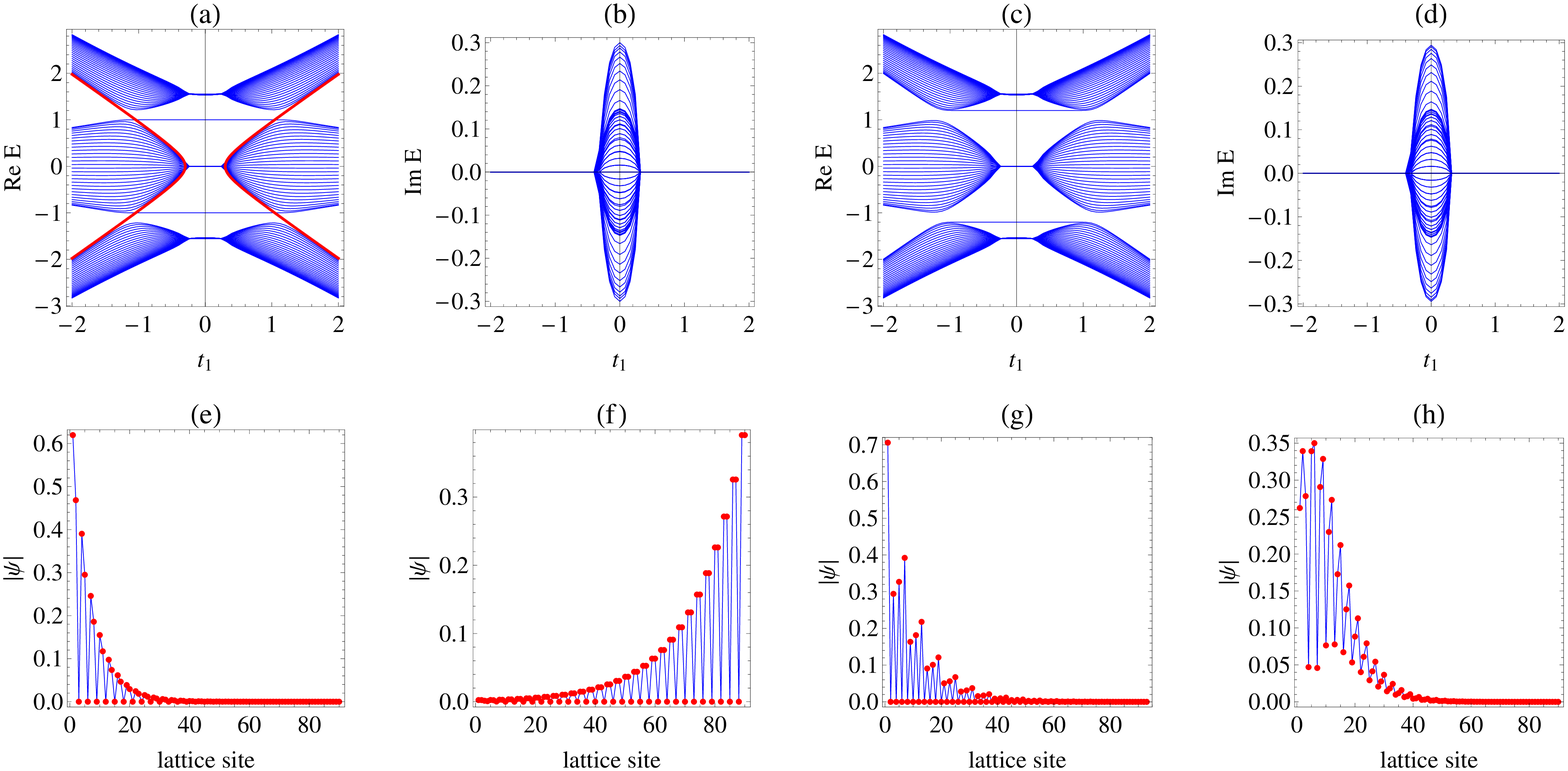}}
\caption{(Top row) The energy spectrum Re$E$ (a,c) and Im$E$ (b,d) of the non-Hermitian SSH3 model~\eqref{H-r} with open boundary condition as a function of $t_1$. Here $t_2=1$, $t_3=1.2$ for (a,b), $t_2=1.2$, $t_3=1$ for (c,d), and $\gamma=0.3$ for all. The type-I localized states are at $E=\pm t_2$ while the type-II localized states with $E=\pm\sqrt{t_1^2-\gamma^2}$ are shown by the thick lines on (a). (Bottom row) The modulus of the wavefunctions of (e) the type-I localized state with $E=t_2$, (f) the type-II localized state with $E=\sqrt{t_1^2-\gamma^2}$, (g) the zero-energy state, and (h) a selected bulk state, respectively. Here $t_2=1$, $t_3=1.2$, $\gamma=0.3$, and $N=90$. $t_1=0.7$ for (e) and $t_1=1.1$ for (f). For (g,h), $t_1=0.7$, $t_2=1.2$, $t_3=1$, and $\gamma=0.2$.  $N=93$ for (g) and $N=90$ for (h).}
\label{p3-all}
\end{figure*}

\subsection{Energy spectrum and symmetry-protected localized states}
The Hamiltonian of the non-Hermitian SSH3 model in real space with finite length is given by
\be
H=\left(
    \begin{array}{cccccc}
      0 & t_1' & & & & \\
      t_1'' & 0 & t_2 & & & \\
       & t_2 & 0 & t_3 &  & \\
       &  & t_3 & \ddots & \ddots &  \\
       &  &  & \ddots & 0 &t_2\\
       &  &  &  & t_2 &0
    \end{array}
  \right). \label{H-r}
\ee
Here $t_1'=t_1+\gamma$ and $t_1''=t_1-\gamma$.
Figure~\ref{p3-all} shows the energy spectrum and profiles of selected eigenstates of the non-Hermitian SSH3 model. 
In the top row, the energy spectrum Re$E$ and Im$E$ are plotted as a function of $t_1$. There are three bands,
two types of localized states with energies $E=\pm\sqrt{t_1^2-\gamma^2}$ and $E=\pm t_2$, and a zero-energy state. We will show that those special states are associated with the symmetries of the non-Hermitian SSH3 model. 

The bottom row of Fig.~\ref{p3-all} shows the modulus of the wavefunctions in real space of the two types of localized states, the zero-energy state, and a typical bulk state, respectively. The non-Hermitian SSH3 model clearly exhibits the non-Hermitian skin effect because the weight of the bulk state tilts towards one end of the system, making it more challenging to distinguish the localized and bulk states. Nevertheless, the two types of localized states and the zero-energy state exhibit distinct profiles, making it possible to distinguish them from the bulk states.

The two types of localized states can be analyzed as follows. We define a diagonal matrix
\be
U=\mbox{diag}\{1,r,r,r,r^2,\cdots,r^M\},\quad r=\sqrt{\frac{t_1-\gamma}{t_1+\gamma}}.
\ee
Then one can verify that $H'=U^{-1}HU$ is a Hermitian matrix with $H'_{3n-2,3n-1}=H'_{3n-1,3n-2}=\sqrt{t_1^2-\gamma^2}$ and the other elements remaining the same as those in $H$. The similar transformation allows us to identify the $E=t_2$ state with the following amplitude
\be
a_{3n-1}&=&a_{3n}=(\frac{1}{r})^{M-n}\Big(-\frac{\sqrt{t_1^2-\gamma^2}}{t_3}\Big)^{M-n} \nonumber \\
&=&\Big(-\frac{t_1+\gamma}{t_3}\Big)^{M-n}, \nonumber \\
a_{3n-2}&=&0.
\ee
The $E=-t_2$ state can be found similarly. The $E=\pm t_2$ states only survive if $t_1 < t_3$.
One the other hand, the $E=t_1$ state becomes a state with $E=\sqrt{t_1^2-\gamma^2}$ and the following amplitude
\be
&&a_{3n-2}=r^{n-1}\Big(-\frac{t_2}{t_3}\Big)^{n-1},
a_{3n-1}=r^n\Big(-\frac{t_2}{t_3}\Big)^{n-1}, a_{3n}=0. \nonumber \\
& &
\ee
Similarly, the $E=-\sqrt{t_1^2-\gamma^2}$ can also be found. The $E=\pm\sqrt{t_1^2-\gamma^2}$ localized states survive if $t_2 < t_3$. Therefore, those symmetry-protected localized states have a zero-amplitude point for every three sites. In contrast, the bulk states do not have such repeated zero-amplitude points. Checking the zero-amplitude points thus differentiates the symmetry-protected localized states from the bulk-states of the non-Hermitian SSH3 model in the presence of the non-Hermitian skin effect.

\begin{figure}
\centerline{\includegraphics[width=\columnwidth]{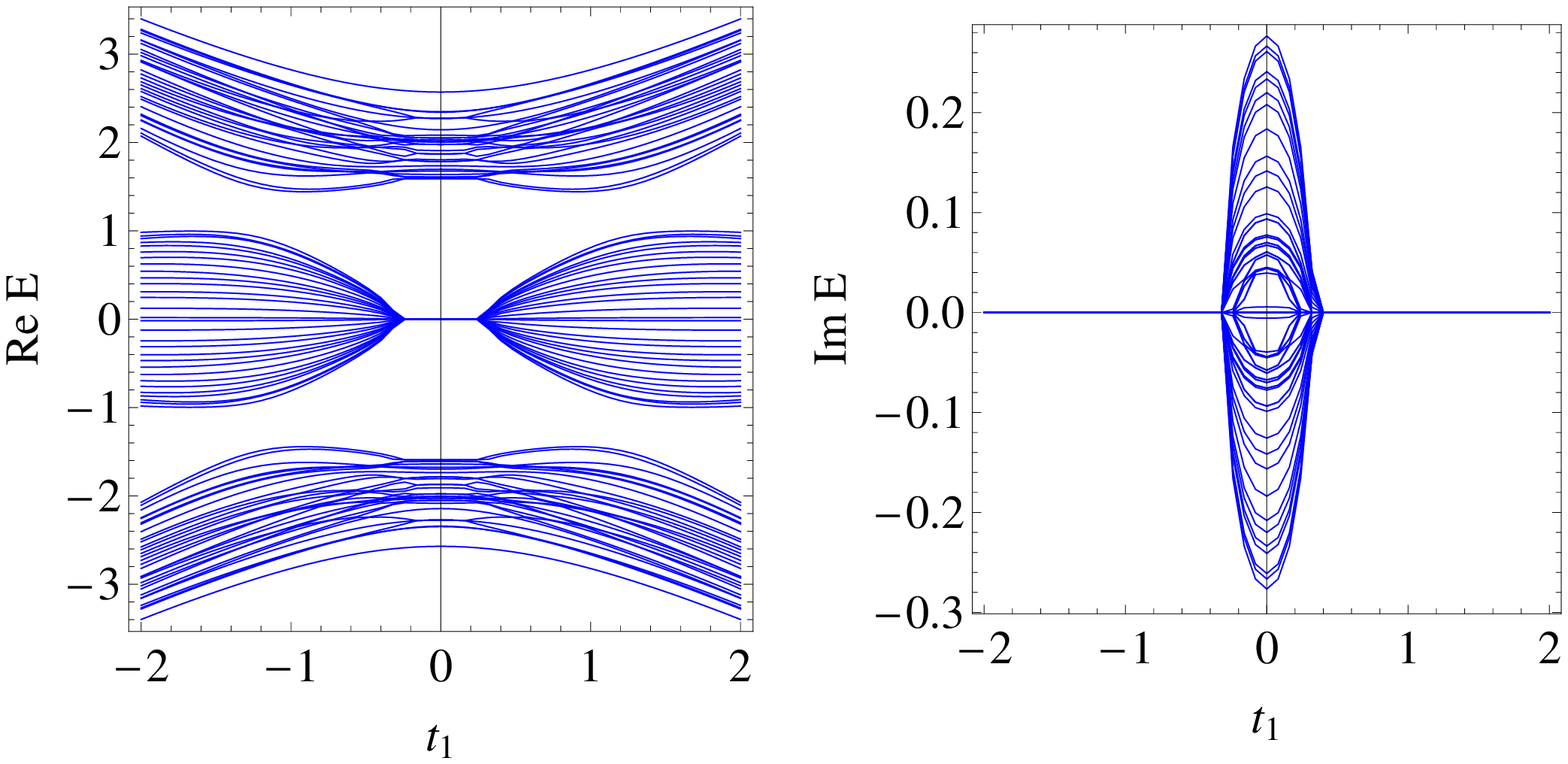}}
\caption{The real part (left panel) and imaginary part (right panel) of the energy spectrum of the non-Hermitian SSH3 model~(\ref{Hr}) with disorder and open boundary condition as a function of $t_1$. In the presence of disorder, the mid-gap states merge into the bulk bands and are no longer visible. Here $t_2=1.2$, $t_3=1$ and $\gamma=0.3$.}
\label{p3-r}
\end{figure}

We want to point out that those symmetry-protected edge states have nothing to do with the winding number shown in Eq.~(\ref{wind}), and they are fragile against disorder. If a random hopping term is added to the SSH3 model, the edge states will merge into the bulk bands while the energy gaps remain open. To illustrate this feature, we introduce the following generalization of the non-Hermitian SSH3 model with disorder:
\be
&&H_r=\sum_{j}\Big[(t_1+\gamma)\dc_{j,1}c_{j,2}+(t_1-\gamma)\dc_{j,2}c_{j,1}\Big]\nonumber\\
&&+\sum_j\Big[t_2(1+r_j)\dc_{j,2}c_{j,3}+t_3\dc_{j,3}c_{j+1,1} + h.c.\Big].\label{Hr}
\ee
Here $r_j$ are independent random variables uniformly distributed between 0 and 1. In Figure \ref{p3-r}, we plot the energy spectrum of $H_r$. One can clear see that the localized states originally located at $E=\pm t_2$ merge into the bulk bands while the bandgaps remain open in the presence of disorder.

For completeness, we also show the wavefunction of the zero-energy state of the non-Hermitian SSH3 model with an odd number of lattice sites as follows.
The zero-energy state is protected by the chiral symmetry, which still applies to the non-Hermitian SSH3 model. The amplitude is given by
\be
&&a_{2n}=0,\quad a_{6m-5}=(-\frac{t_1-\gamma}{t_1})^{m-1},\\
&&a_{6m-3}=-\frac{t_1-\gamma}{t_2}(-\frac{t_1-\gamma}{t_1})^{m-1},\\
&&a_{6m-1}-\frac{t_3}{t_2}(-\frac{t_1-\gamma}{t_1})^{m}.
\ee
Here $n=1,\cdots,(N-1)/2$ and $m=1,\cdots (N-1)/6$. Therefore, the amplitude vanishes on every other site, so the profile of the zero-energy state can be distinguished from those of the bulk states.

\begin{figure}
\centerline{\includegraphics[width=0.6\columnwidth]{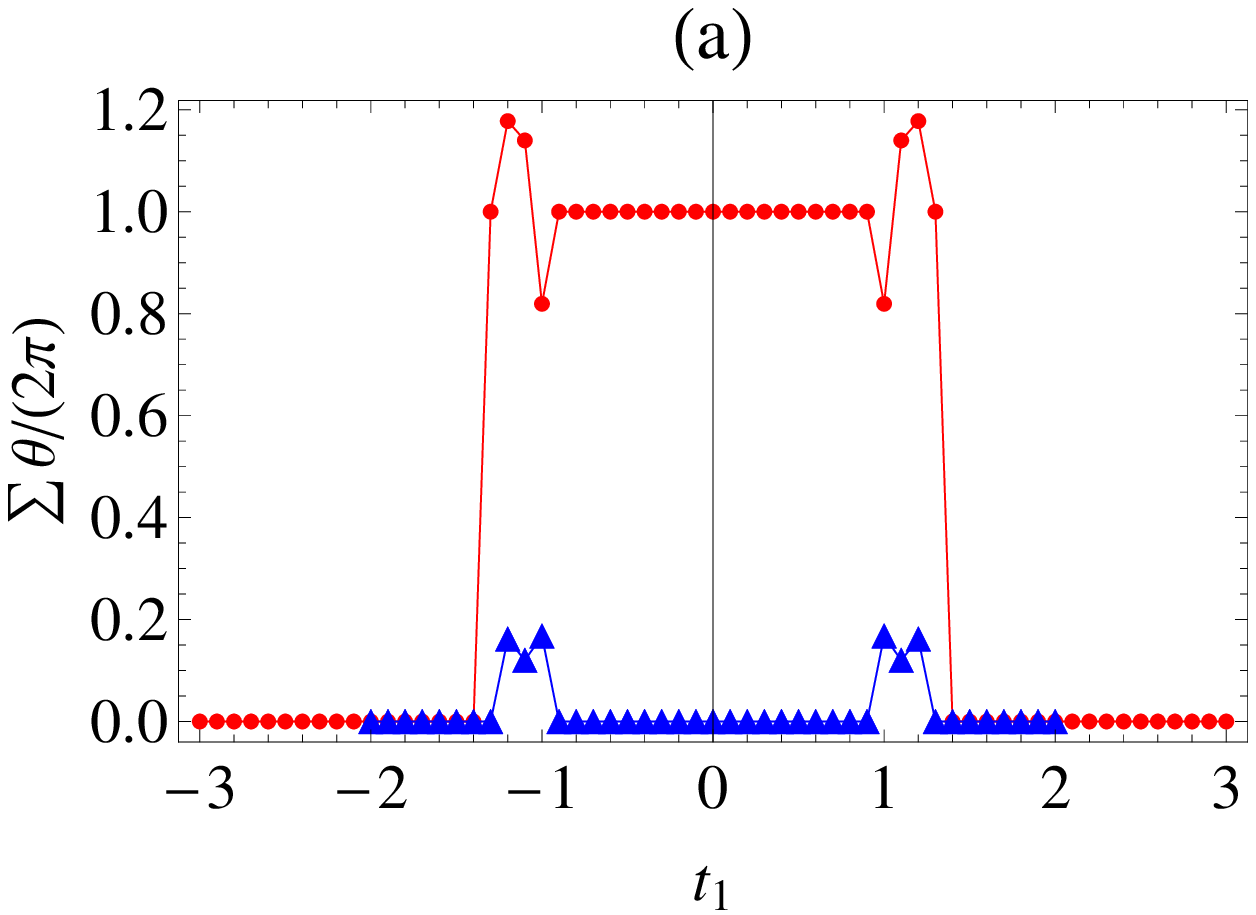}}
\caption{The total Zak phase $\sum_{n=1}^3\theta_n$ of the non-Hermitian SSH3 model as a function of $t_1$. The circles (triangles) correspond to the case with $t_2=1$ and $t_3=1.2$ ($t_2=1.2$, $t_3=1$). $\gamma=0.3$.}
\label{Zak}
\end{figure}

\subsection{Zak phase and coexistence of localized states}
Since the winding number~\eqref{wind} does not associate directly with the localized states in finite systems, here we consider another quantity, known as the Zak phase~\cite{Zak89}, that may reflect the band-topology. For the non-Hermitian SSH3 model, the Zak phase of the $n$-th band is obtained from
\be
\theta_n=-i\int_0^{2\pi}dk\ep{u^L_n|\frac{d}{dk}|u^R_n},\quad n=1,2,3.
\ee
Here $\ket{u^R_n}$ and $\ket{u^L_n}$ are the $n$-th eigenvectors of $H$ and $H^{\dag}$, respectively. The behavior of the Zak phase is similar for the Hermitian and non-Hermitian SSH3 models. In the following, we analyze the Zak phase of the non-Hermitian SSH3 model. For a selected band, the Zak phase of the SSH3 model is usually nonzero but also not quantized, suggesting no topological meaning. According to Ref.~\cite{Chen2018}, however, the sum of the Zak phase of all the bands, $\sum_{n=1}^3\theta_n$, may be related to the band topology. We contrast the behavior of $\sum_{n=1}^3\theta_n$ in Figure \ref{Zak} with different values of $t_2$ and $t_3$, showing the dependence on the hopping coefficients. The regime with non-vanishing total Zak phase can be mapped out, and we found
\be
\frac{1}{2\pi}\sum_{n=1}^3\theta_n=\left\{
                       \begin{array}{ll}
                         1, & \,|t_1| \& |t_2| < t_3, \\
                         0, & \,\mbox{otherwise.}
                       \end{array}
                     \right.
\ee
Therefore, the total Zak phase of the SSH3 model is quantized.

Moreover, the coexistence regime of the type-I and type-II localized states coincides with that of the nonzero total Zak phase of the SSH3 model. In this respective, the total Zak phase may serve as an indicator of where the two types of localized states can be observed in the non-Hermitian SSH3 model. Explicitly, if $\sum_n \theta_n/(2\pi)$ is finite, there are two (positive-energy) localized states on the two ends of the chain. When $\sum_n \theta_n/(2\pi)=0$, however, there may be one (positive-energy) localized state or none.

\section{SSH4 model}\label{sec:SSH4}
\subsection{Hermitian model}
After considering the SSH3 generalizations, we move on to the Hermitian model with four lattice-sites in one unit cell, known as the SSH4 model \cite{SSH4}. With periodic boundary condition, the Bloch Hamiltonian in momentum space is
\be
H=\left(
    \begin{array}{cccc}
      0 & t_1 & 0 & t_4e^{-ik} \\
      t_1 & 0 & t_2 & 0 \\
      0 & t_2 & 0 & t_3 \\
      t_4e^{ik} & 0 & t_3 & 0
    \end{array}
  \right).
\ee
The eigenvalues are given by
\be
&&E=\pm\Big(\frac{B\pm\sqrt{B^2-4C}}{2}\Big)^{1/2},\quad B=t_1^2+t_2^2+t_3^2+t_4^2, \nonumber \\
&&C=(t_1t_3)^2+(t_2t_4)^2-2t_1t_2t_3t_4\cos k.
\ee
There are four bands. If $t_1t_3=\pm t_2t_4$, the gap between the middle two bands is closed at $k=0$ or $k=\pi$.

We briefly summarize the symmetry of the SSH4 model. By introducing the matrix $\Gamma_4 =I_2\otimes\sigma_z$ with $I_2$ being the 2 by 2 identity matrix, one can verify that
\be
\Gamma_4 H+H\Gamma_4=0.
\ee
Therefore, the SSH4 model has a chiral symmetry and belongs to the AIII class. In 1D, the AIII class is topologically nontrivial~\cite{Chiu2016}, and its topology is captured by the winding number, which will be defined shortly. Thus, the Hermitian SSH4 model is already topological and quite different from the SSH3 model.
We introduce the following matrix
\be
S=\left(
    \begin{array}{cccc}
      1 & 0 & 0 & 0 \\
      0 & 0 & 1 & 0 \\
      0 & 1 & 0 & 0 \\
      0 & 0 & 0 & 1
    \end{array}
  \right).
\ee
Then,
\be
SHS^{-1}=\left(
           \begin{array}{cc}
             0 & g \\
             g^{\dagger} & 0 \\
           \end{array}
         \right),\quad
g=\left(
           \begin{array}{cc}
             t_1 & t_4 e^{-ik} \\
             t_2 & t_3 \\
           \end{array}
         \right).
\ee
The winding number can be obtained as follows~\cite{Chiu2016}.
\be
W=\frac{1}{2\pi i}\int_0^{2\pi}dk \frac{d\ln(\det g)}{dk}
=\frac{1}{2\pi i}\int_0^{2\pi}dk\, z^{-1}\frac{dz}{dk}.
\ee
Here $\phi$ is the polar angle of the complex number $z=\det (g)=t_1t_3-t_2t_4e^{-ik}$. If $|t_1t_3|<|t_2t_4|$, the path of the integration of $z$ will circle around the point $z=0$ once. One the other hand, if $|t_1t_3|<|t_2t_4|$, the path of the integration of $z$ does not enclose $z=0$. Therefore,
\be
W=\left\{
    \begin{array}{ll}
      1, & |t_1t_3|<|t_2t_4|; \\
      0, & |t_1t_3|>|t_2t_4|.
    \end{array}
  \right.
\ee

\subsubsection{Symmetry-protected localized states}
For the SSH4 model with open boundary condition, one can start with the wavefunction $(\cdots, a_{4n}, a_{4n+1}, a_{4n+2}, a_{4n+3}, \cdots)^T$ and obtain the following eigenvalue equations:
\be
&&t_1 a_{4n-3}-E a_{4n-2}+t_2 a_{4n-1}=0,\\
&&t_2 a_{4n-2}-E a_{4n-1}+t_3 a_{4n}=0,\\
&&t_3 a_{4n-1}-E a_{4n}+t_4 a_{4n+1}=0,\\
&&t_4 a_{4n}-E a_{4n+1}+t_1 a_{4n+2}=0.
\ee
with $n=1,\cdots,M$ and $M=N/4$. The open boundary condition imposes the constraints:
\be
&&-Ea_1+t_1 a_2=0,\\
&&t_3a_{4M-1}-E a_{4M}=0.
\ee
There is a pair of localized states associated with a symmetry of the SSH4 model, which will be analyzed here.

For the state with $E=\sqrt{t_2^2+t_3^2}$, we find
\be
&&a_{4n-3}=0,\quad
a_{4n-2}=t_2\Big(-\frac{t_1}{t_4}\Big)^{M-n},\\
&&a_{4n-1}=\sqrt{t_2^2+t_3^2}\Big(-\frac{t_1}{t_4}\Big)^{M-n},\quad
a_{4n}=t_3\Big(-\frac{t_1}{t_4}\Big)^{M-n}.
\ee
The state is normalizable if $t_1<t_4$, and it is localized at the right end of the chain. The boundary condition $a_2=0$ is satisfied in the thermodynamic limit and is approximately satisfied in a finite chain. One can follow a similar derivation to obtain the localized state with $E=-\sqrt{t_2^2+t_3^2}$.
The localized states with $E=\pm\sqrt{t_2^2+t_3^2}$ are due to a symmetry of the SSH4 model. We introduce a diagonal matrix $\Gamma'_4=\mbox{diag}\{-1,1,1,1-1,\cdots,1\}$. Then it follows
\be
&&\Gamma'_4 H\Gamma'_4 +H=2H_0,\\
&&H_0=\mbox{diag}\{0,A,\cdots,A\},\quad
A=\left(
      \begin{array}{ccc}
        0 & t_2 & 0 \\
        t_2 & 0 & t_3 \\
        0 & t_3 & 0 \\
      \end{array}
    \right).\nonumber
\ee
Thus, one can find an eigenstate $\psi$ satisfying $\Gamma'_4 \psi=\psi$ and $H_0\psi=\sqrt{t_2^2+t_3^2}\psi$. If $\psi$ is also an eigenstate of $H$, the eigenvalue of $H$ must be $\sqrt{t_2^2+t_3^2}$. Similarly, the symmetry also leads to the $E=-\sqrt{t_2^2+t_3^2}$ state. Therefore, the states are not associated with the band-topology.

\begin{figure}
\centerline{\includegraphics[width=\columnwidth]{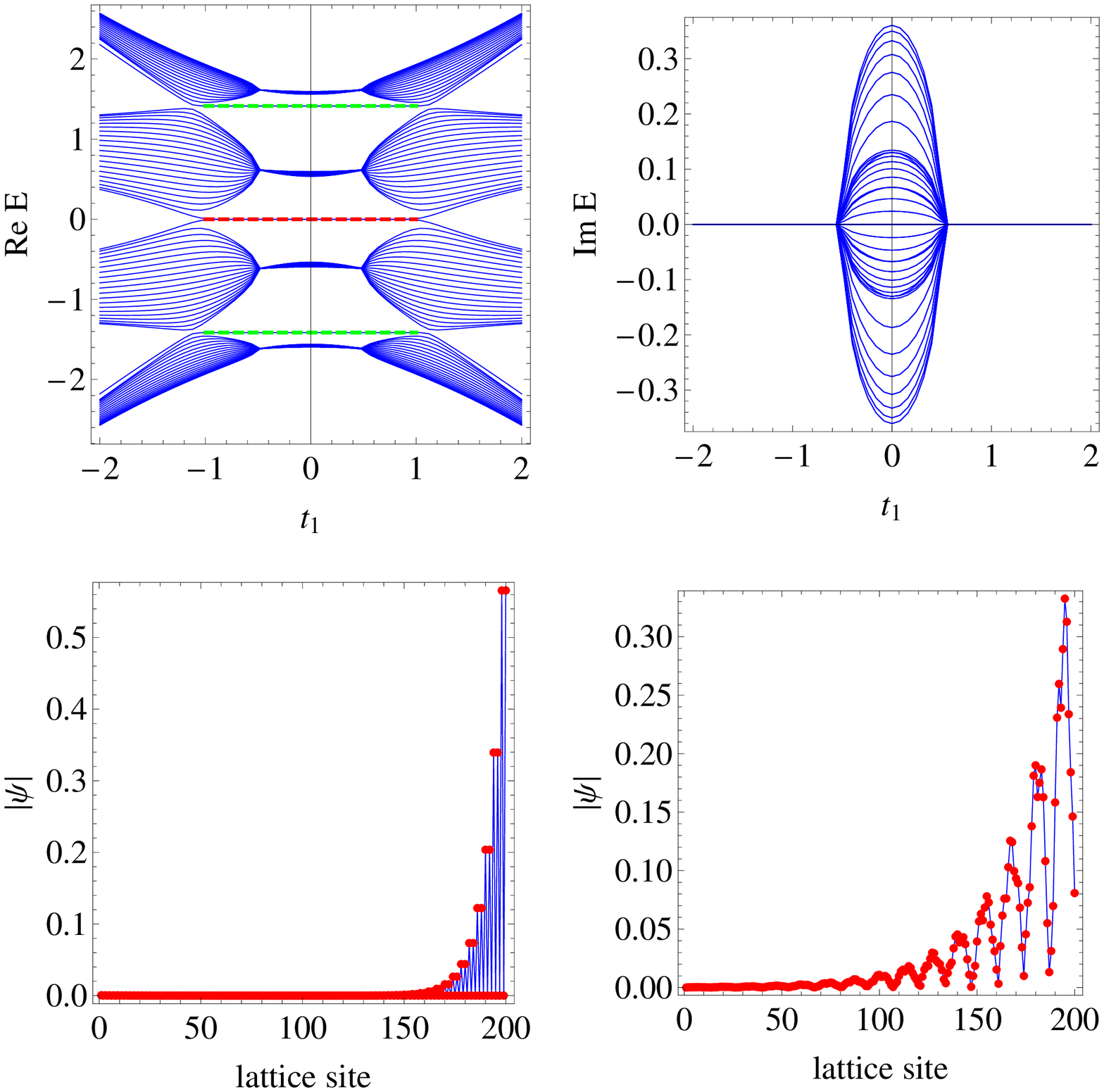}}
\caption{(Top row) The energy spectrum Re$E$ (left) and Im$E$ (right) of the non-Hermitian SSH4 model with open boundary condition as a function of $t_1$. Here $t_2=t_3=t_4=1$ and $\gamma=0.5$. The localized states associated with the winding number and the special symmetry are emphasized by thick dashed lines. (Bottom row) The modulus of the wavefunctions of the zero-energy edge state (left) and a bulk state (right), respectively. We assume $t_1=0.7$, $t_2=t_3=t_4=1$, $\gamma=0.1$ and $N=200$.}
\label{psi}
\end{figure}

\subsection{Non-Hermitian generalization}
\begin{figure}
\centerline{\includegraphics[width=0.6\columnwidth]{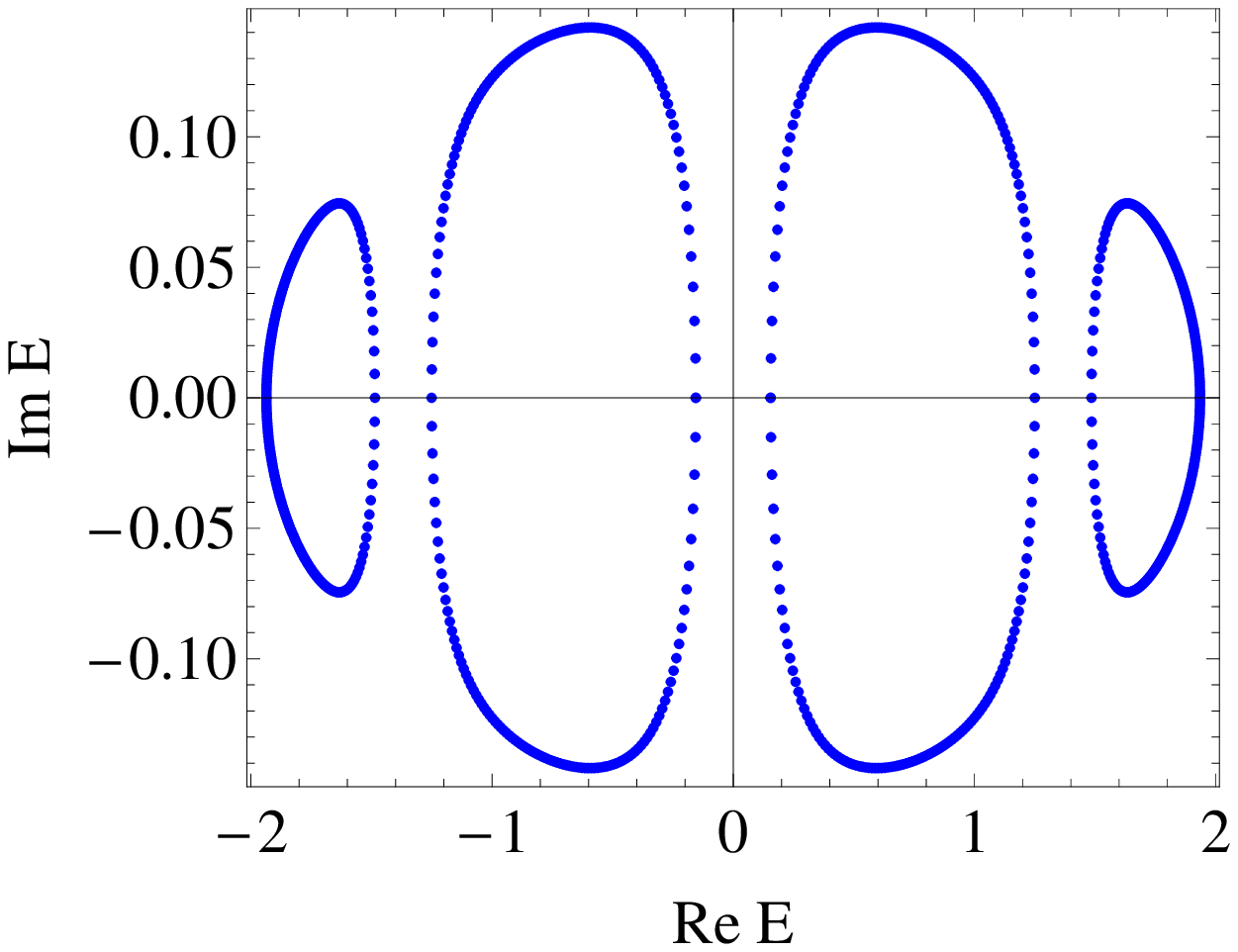}}
\caption{Energy spectrum of the non-Hermitian SSH4 model ~\eqref{Hk-4} on the complex plane. Here $t_1=1.2$, $t_2=t_3=1$, $t_4=0.7$, $\gamma=0.4$, and periodic boundary condition has been used.}
\label{E-SSH4}
\end{figure}

We consider a non-Hermitian generalization of the SSH4 model with periodic boundary condition given by the Bloch Hamiltonian
\be
H=\left(
    \begin{array}{cccc}
      0 & t_1+\gamma & 0 & t_4e^{-ik} \\
      t_1-\gamma & 0 & t_2 & 0 \\
      0 & t_2 & 0 & t_3 \\
      t_4e^{ik} & 0 & t_3 & 0
    \end{array}
  \right).\label{Hk-4}
\ee
Here the non-Hermitian property is introduced by letting the hopping between site 1 and site 2 be non-reciprocal. The eigenvalues are
\be
E&=&\pm\Big(\frac{B\pm\sqrt{B^2-4C}}{2}\Big)^{1/2},~ B=t_1^2+t_2^2+t_3^2+t_4^2-\gamma^2, \nonumber \\
C&=&(t_1t_3)^2+(t_2t_4)^2-\gamma^2t_3^2-2t_1t_2t_3t_4\cos k \nonumber \\
& &-2i\gamma t_2t_3t_4\sin k.
\ee
The gap between the middle two bands will close if $C=0$. The condition $\textrm{Im}C=0$ requires $k=0$ or $k=\pm\pi$. Then the condition for closing the gap is
\be
(t_1t_3\pm t_2t_4)^2=\gamma^2t_3^2\label{t-ga}.
\ee
The non-Hermitian SSH4 model has a line gap on the complex plane as illustrated in Fig.~\ref{E-SSH4}, different from the point gap of the non-Hermitian SSH3 model shown in Fig.~\ref{E-C}.
In the non-Hermitian model, the gap closing point obtained from the system with periodic boundary condition is generally different from the point where the edge states emerge in the system with open boundary condition. This is because the bulk spectrum with open boundary can be different from the spectrum computed in momentum space, leading to the non-Hermitian skin effect~\cite{TorrePRB18,Wang1,Wang2}.

Here we demonstrate the non-Hermitian skin effect of the SSH4 model.  According to Eq.~(\ref{t-ga}), the gap of the SSH4 model in momentum space closes at $t_1=\pm(\dfrac{t_2t_4}{t_3}+\gamma)$. With the selected values $t_2=t_3=t_4=1$ and $\gamma=0.5$, the topological transition with periodic boundary condition  should occur at $t_1=\pm1.5$. The corresponding real-space Hamiltonian with open boundary condition is
\be
H=\left(
    \begin{array}{ccccccc}
      0 & t_1' & & & & &\\
      t_1'' & 0 & t_2 & & & &\\
       & t_2 & 0 & t_3 &  & &\\
       &  & t_3 & 0 & t_4 & & \\
       &  &  & t_4 & \ddots & \ddots& \\
       &  &  &  & \ddots & 0 &t_3\\
       &  &  &  &  & t_3 &0
    \end{array}
  \right). \label{H-r4}
\ee
Here $t_1'=t_1+\gamma$ and $t_1''=t_1-\gamma$. In the upper panels of Figure \ref{psi}, we plot the real and imaginary parts of the energy spectrum of the non-Hermitian SSH4 model with open boundary condition. The grid size is $N=20$, $t_2=t_3=t_4=1$, and $\gamma=0.5$. One can see in the upper left panel of Figure \ref{psi} the edge state appears at $t_1=\pm\sqrt{(\frac{t_2t_4}{t_3})^2+\gamma^2}\approx\pm1.1$, different from the transition point estimated from the momentum-space calculation.
The chiral symmetry is the reason behind the zero-energy edge state.

The transition point of the real-space calculation can be explained as follows. We introduce a diagonal matrix
\be
U=\mbox{diag}\{1,r,r,r,r,r^2,\cdots,r^N\},\quad r=\sqrt{\frac{t_1-\gamma}{t_1+\gamma}}.
\ee
Then a similar transformation of $H$ by $U$ leads to a Hermitian matrix
\be
H'=U^{-1}HU=\left(
    \begin{array}{ccccccc}
      0 & t_u & & & & &\\
      t_u & 0 & t_2 & & & &\\
       & t_2 & 0 & t_3 &  & &\\
       &  & t_3 & 0 & t_4 & & \\
       &  &  & t_4 & \ddots & \ddots& \\
       &  &  &  & \ddots & 0 &t_3\\
       &  &  &  &  & t_3 &0
    \end{array}
  \right)
\ee
with $t_u=\sqrt{t_1^2-\gamma^2}$. Importantly, $H'$ is nothing but the Hermitian SSH4 model. Therefore, the transition point is given by $t_ut_3=t_2t_4$, or equivalently $t_1=\pm\sqrt{(\frac{t_2t_4}{t_3})^2+\gamma^2}$, according to the Hermitian model. We will first present a more detailed analysis of the edge states and then address how to reconcile the results of the non-Hermitian model with different boundary conditions.

In the lower panels of Figure \ref{psi}, we present the results of the edge and bulk states to contrast their difference. In the lower left and right panels, we plot the modulus of the wavefunctions of the edge state and a typical bulk state, respectively, as a function of the lattice index. Here we assume $t_1=0.7$, $t_2=t_3=t_4=1$, $\gamma=0.1$, and use the grid size $N=200$. While the topological edge state with $E=0$ clearly localizes at one end of the system, the bulk state shows similar localization behavior due to the non-Hermitian skin effect.

Nevertheless, the topological edge state has a particular feature, allowing it to be distinguished from the bulk states. The zero-energy edge state has an analytic expression as follows. From the eigenvalue equation $H\psi=0$ with $\psi=(a_1,\dots,a_N)^T$ for the zero-energy state, we find the following relations:
\be
&&(t_1-\gamma)a_{4n-3}+t_2 a_{4n-1}=0,\\
&&t_2 a_{4n-2}+t_3 a_{4n}=0,\\
&&t_3 a_{4n-1}+t_4 a_{4n+1}=0,\\
&&t_4 a_{4n}+(t_1+\gamma) a_{4n+2}=0.
\ee
The open boundary condition is enforced by $(t_1+\gamma)a_2=0$ and $t_3a_{N-1}=0$. By assuming $a_{N-1}=0$ and $a_N=1$, we find the solution
\be
&&a_{4n-2}=\Big(\frac{t_3}{t_2}\Big)^{M-n+1}\Big(\frac{t_1+\gamma}{t_4}\Big)^{M-n},\\
&&a_{4n}=\Big(\frac{t_3}{t_2}\Big)^{M-n}\Big(\frac{t_1+\gamma}{t_4}\Big)^{M-n},\\
&&a_{4n-1}=a_{4n-3}=0.
\ee
Here $n=1,\cdots,M$ and $M=N/4$. If we assume $t_1+\gamma<t_4$ and $t_3<t_2$, then $a_{2n}$ decreases exponentially as $n$ becomes small. The boundary condition $a_2=0$ is then approximately satisfied in a long chain. We thus obtain an approximate wavefunction of the edge states. Importantly, the amplitude of the zero-energy edge state should virtually vanish on every other site, as one can see in Fig.~\ref{psi}. The bulk states, in contrast, do not have such a pattern, so one can differentiate the edge state by its wavefunction profile.

Incidentally, the symmetry-protected localized states also survives in the non-Hermitian SSH4 model. From the Hamiltonian~\eqref{H-r4}, the localized state at $E=\sqrt{t_2^2+t_3^2}$ has the profile
\be
&&a_{4n-3}=0,\quad
a_{4n-2}=t_2\Big(-\frac{t_1+\gamma}{t_4}\Big)^{M-n},\\
&&a_{4n-1}=\sqrt{t_2^2+t_3^2}\Big(-\frac{t_1+\gamma}{t_4}\Big)^{M-n},\\
&&a_{4n}=t_3\Big(-\frac{t_1+\gamma}{t_4}\Big)^{M-n}.
\ee
The localized state survives in the regime with $t_1+\gamma < t_4$.
Similarly, the localized state with $E=-\sqrt{t_2^2+t_3^2}$ can be found in the same regime. The two localized states correspond to the two flat lines at $Re(E)\sim 1.5$ on the upper-left corner of Fig.~\ref{psi}. We remark that those localized states are associated with the symmetry, not the band-topology.

\subsection{Restoration of bulk-boundary correspondence}
The bulk-boundary correspondence of non-Hermitian topological systems may be restored in many ways~\cite{TorreJPhys20}.
After understanding the zero-energy edge state better, we will show that the bulk-boundary correspondence of the non-Hermitian SSH4 model can be restored by two different methods. In the first method, we introduce a generalized Bloch Hamiltonian~\cite{Yokomizo19} by making the substitution $e^{ik}\to\beta$ in the momentum-space model (\ref{Hk-4}). The result is
\be
H(\beta)=\left(
    \begin{array}{cccc}
      0 & t_1+\gamma & 0 & t_4\beta^{-1} \\
      t_1-\gamma & 0 & t_2 & 0 \\
      0 & t_2 & 0 & t_3 \\
      t_4\beta & 0 & t_3 & 0
    \end{array}
  \right).
\ee
Here $\beta$ is a complex number. Because of the non-Hermitian skin effect, we usually have $|\beta|\neq1$. The trajectory of $\beta$ in the complex plane is a closed loop, which is called the generalized Brillouin zone \cite{Wang1}, denoted by $C_\beta$. It is a generalization of the unit circle described by the factor $e^{ik}$ on the complex plane. With the introduction of the generalized Bloch method, the energy eigenvalues can be obtained from
\be
&&E^4-BE^2+C(\beta)=0\\
&&\mbox{with}\quad B=t_1^2+t_2^2+t_3^2+t_4^2-\gamma^2, \nonumber\\
&&C(\beta)=(t_1t_3)^2+(t_2t_4)^2-\gamma^2t_3^2 \nonumber \\
&&-t_1t_2t_3t_4(\beta+\beta^{-1})-\gamma t_2t_3t_4(\beta-\beta^{-1}).\nonumber
\ee
This is a quadratic equation of $\beta$, which has two complex roots $\beta_1$ and $\beta_2$. The continuum band can be obtained by requiring $|\beta_1|=|\beta_2|$. To determine $\beta$, we notice that the same $E$ corresponds to both $\beta_1$ and $\beta_2$, and we also have $\beta_1=\beta_2e^{i\theta}$ for some phase angle $\theta$. These considerations lead to the following equation
\be
C(\beta)=C(\beta e^{i\theta}),
\ee
which can determine $\beta$ for a fixed $\theta$. One can see that this equation simplifies to
\be
&&(t_1+\gamma)\beta+(t_1-\gamma)\beta^{-1}\nonumber\\
&&\qquad =(t_1+\gamma)\beta e^{i\theta}+(t_1-\gamma)\beta^{-1}e^{-i\theta},
\ee
and its solution is $|\beta|=\sqrt{\dfrac{t_1-\gamma}{t_1+\gamma}}$. Therefore, $C_{\beta}$ is still a circle on the complex plane but with a radius smaller than $1$.

To calculate the winding number, we again transform $H(\beta)$ to an off-diagonal form
\be
&&SH(\beta)S^{-1}=\left(
           \begin{array}{cc}
             0 & g_1 \\
             g_2 & 0 \\
           \end{array}
         \right),\\
&&g_1=\left(
           \begin{array}{cc}
             t_1+\gamma & t_4\beta^{-1} \\
             t_2 & t_3 \\
           \end{array}
         \right),\quad
g_2=\left(
           \begin{array}{cc}
             t_1-\gamma & t_2 \\
             t_4\beta & t_3 \\
           \end{array}
         \right).\nonumber
\ee
With the introduction of $\beta$, the winding number can be generalized to
\be
W=\frac{1}{2\pi i}\int_0^{2\pi}d\theta\, z^{-1}\frac{dz}{d\theta}.
\ee
Here $z=\det (g_1)=(t_1+\gamma)t_3-t_2t_4|\beta|^{-1}e^{-i\theta}$. Then it can be shown that
\be
W=\left\{
    \begin{array}{ll}
      1, & t_1<\sqrt{(\frac{t_2t_4}{t_3})^2+\gamma^2}; \\
      0, & t_1>\sqrt{(\frac{t_2t_4}{t_3})^2+\gamma^2}.
    \end{array}
  \right.
\ee
Therefore, the regime of nonzero winding number agrees with the appearance of the zero-energy edge states computed from the case with open boundary condition. Since the non-Hermitian SSH4 model has a line gap, the results for systems with point gaps in Ref.~\cite{OkumaPRL20} may not apply.

In the second method, we directly calculate the winding number in real space with open boundary condition, following Refs.~\cite{Wang19}. To achieve the goal, we need to obtain the left- and right- eigenvectors of Eq.~(\ref{H-r4}), given by
\be
H\ket{u_n^R}=E_n\ket{u_n^R},\quad H^{\dagger}\ket{u_n^L}=E^*_n\ket{u_n^L}.
\label{bi-otho}
\ee
Those eigenvectors allow us to introduce the following $Q$-matrix:
\be
Q=\sum_{\textrm{Re}E_n>0}\ket{u_n^R}\bra{u_n^L}-\sum_{\textrm{Re}E_n<0}\ket{u_n^R}\bra{u_n^L},
\ee
satisfying $Q^2=I$, where $I$ is the identity operator.
The real-space winding number \cite{Kitaev06,Prodan,Wang19} is then given by
\be
W=\frac{1}{2L}\mbox{Tr}'\Big(\Gamma_NQ[Q,X]\Big).
\label{rwind}
\ee
Here $\Gamma_N=I_N\otimes\Gamma$ with $I_N$ being the $N$ by $N$ identity matrix and $\Gamma=I_2\times\sigma_3$. We also define $X=X_1\otimes I_4$ with the position operator $(X_1)_{ij}=i\delta_{ij}$. To eliminate the boundary effects, we divide the 1D system into three segments with lengths $l$, $L$, and $l$, satisfying $L+2l=N$. The partial trace Tr$'$ means the summation is only over the middle segment $L$. The real-space winding number $W$ was first proposed by Kitaev~\cite{Kitaev06} for a Hermitian 1D model with chiral symmetry. It has been shown to be the same as the momentum-space $W$ if the grid size is large enough~\cite{Kitaev06}. One should assume a large enough $l$ in order to avoid the boundary effect, but there is no other constraint such as $l\ll L$. If calculated correctly, $W$ should be real and quantized. However, the eigenvectors of the non-Hermitian model are calculated numerically and may be subjected to numerical errors, which in turn may cause a slight deviation of $W$ from its quantized values.

\begin{figure}
\centerline{\includegraphics[width=0.6\columnwidth]{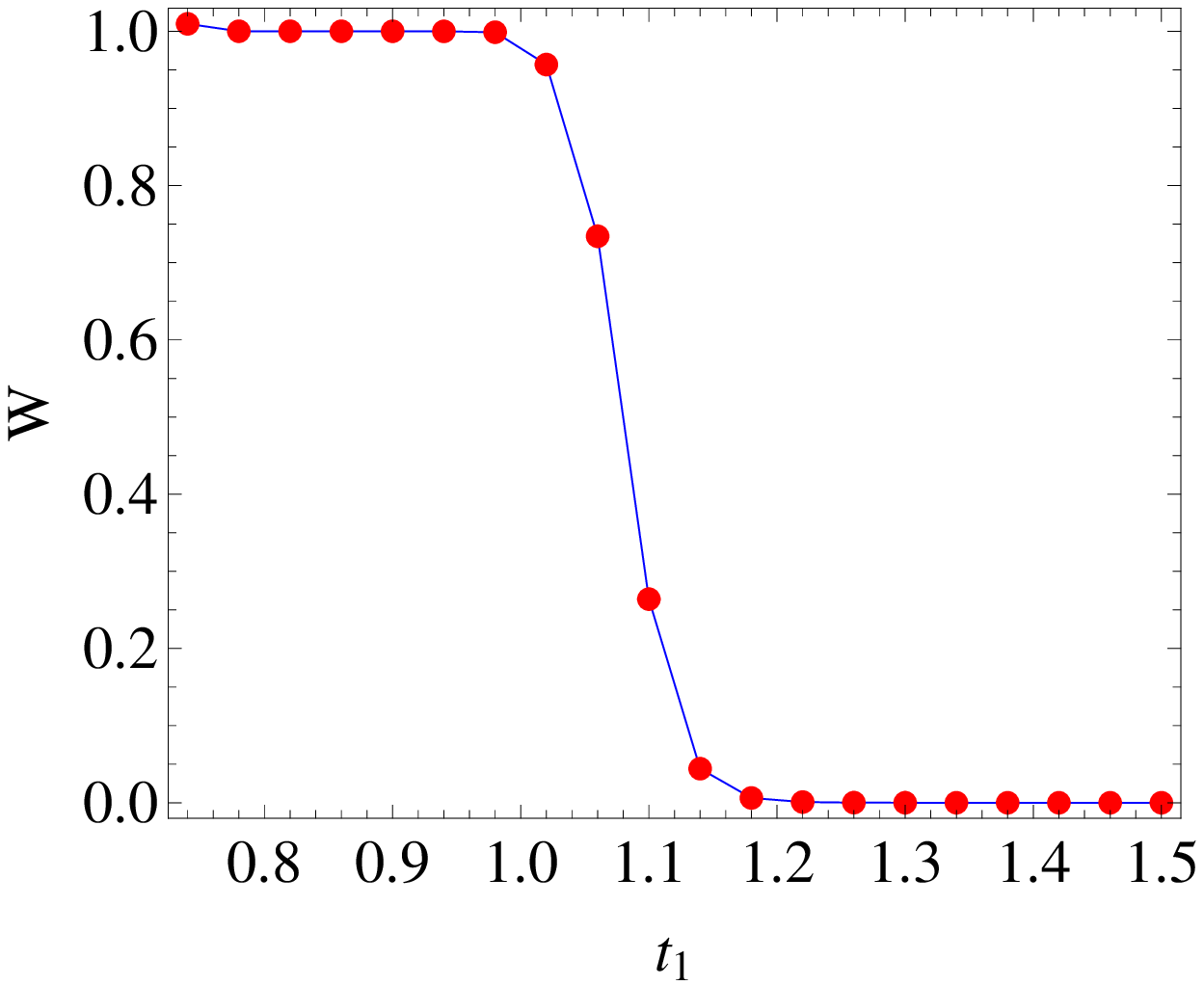}}
\caption{The real-space winding number, shown in Eq.~(\ref{rwind}), of the non-Hermitian SSH4 model as a function of $t_1$. Here $t_2=t_3=t_4=1$ and $\gamma=0.5$.}
\label{W-t1}
\end{figure}

Figure \ref{W-t1} shows the real-space winding number of Eq.~(\ref{rwind}) as a function of $t_1$ with $t_2=t_3=t_4=1$ and $\gamma=0.5$. The grid size is $N=50$ and we use $l=10$. There is a transition of the values of $W$ around $t_1\approx1.1$, which agrees with the location where the edge state emerges. Therefore, the bulk-boundary correspondence is restored for the non-Hermitian SSH4 model if the real-space $W$ is considered.

\subsection{Zak phase}
For the non-Hermitian SSH4 model, the winding number may be inferred from the Zak phase~\cite{Zak89}, which is a measurable quantity used to identify the topology of the Hermitian SSH model~\cite{ZakPhase} in cold atoms. The Zak phase can be obtained from a line integral of the Berry connection. Since the non-Hermitian SSH4 model has the bi-orthogonal eigenstates defined in Eq.~(\ref{bi-otho}), we introduce the non-Abelian Berry connections as
\be
A_{mn}(k)=-i\ep{u^L_m|\frac{\p}{\p k}|u^R_n}.
\ee
Here $m,n$ take the indices of the two lowest energy states for a half-filled system, and the Berry connection defined above is a 2 by 2 matrix. In terms of this Berry connection, the Zak phase after the system traverses the Brillouin zone is introduced as
\be
\theta=\oint dk\,\mbox{Tr} A(k).
\ee
Similar to the Hermitian case, there is a simple relation between the Zak phase and the winding number, given by  $\theta=\pi W~(\mbox{mod}\,2\pi)$. Explicitly, $\theta=0$ if the winding number is even and $\theta=\pi$ if the winding number is odd. In contrast to the non-Hermitian SSH3 model, the sum of the Zak phase of all the four bands of the SSH4 model is always zero.
The Berry connection may also be thought of as the expectation value of the position operator because Tr$A=\sum_{n=1}^2\ep{u^L_n|x|u^R_n}$. Therefore, the Zak phase defined above also reflect the total polarization of the system with the lower two bands filled if the system is charged. We remark that the Zak phase of non-Hermitian models involves both the left and right eigenvectors, so its measurements may be challenging.

\section{Experimental implications}\label{sec:exp}
The SSH model has been realized in experiments using ultracold atoms in optical potentials forming a 1D superlattice~\cite{ZakPhase}. The Zak phase has been detected, showing a quantized difference between different topological regimes.
The Hermitian SSH4 model has also been realized in cold-atom experiments ~\cite{Xi19}. The Hermitian SSH3 model has been proposed to be realizable using cold-atoms~\cite{HePRA18} as well. By coupling additional atoms in augmented optical potentials, non-Hermitian effects may be introduced to cold-atoms systems via the reservoir effects~\cite{Gou20}. Given the rapid developments of trapping and manipulating cold-atoms, the non-Hermitian generalizations of the SSH model may be realized by cold-atom quantum simulators. Other quantum simulators, for example those use photonics~\cite{ZhuPhotonics20,Weidemann20,XiaoNatPhys20}, may also be suitable for demonstrating non-Hermitian behavior. We emphasize that the non-Hermitian SSH3 and SSH4 models are relatively simple and may be accurately analyzed with the help of quantum simulators. Nevertheless, we have shown that the two models cover many signature phenomena of non-Hermitian physics.

On the other hand, it has been demonstrated that one may use classical electric circuits to mimic the behavior of topological systems~\cite{LeeTopolectric18}, and the circuit-analogue of the SSH model has been realized. It is also possible to introduce non-Hermitian effects by engineering the circuit simulators~\cite{Helbig20}. The non-Hermitian skin effect has been demonstrated recently~\cite{Hofmann20}. Therefore, the non-Hermitian generalizations of the SSH model may also find their realizations in electric-circuit simulators. One advantage of using the cold-atom or circuit simulators to study the non-Hermitian effects discussed here is the broad tunability of the parameters in the simulators, which will allow a systematic verification of the phenomena without distractions from irrelevant material properties.

We mention that the dynamics of the Hermitian SSH model can lead to quantized transport, which also has been demonstrated in cold-atom systems~\cite{Nakajima2016,Lohse2016}. Moreover, the topological edge states of the SSH model are proposed to cause quantum memory effects in boundary-induced dynamics~\cite{He2016a}. Therefore, future research integrating quantum dynamics and non-Hermitian effects is expected to unveil more exciting dynamical phenomena.

\section{Conclusion}\label{sec:conclusion}
By combining the generalizations of higher periods of the hopping coefficients and non-Hermitian effect, we have presented interesting physics of the non-Hermitian SSH3 and SSH4 models. While the Hermitian SSH3 model does not have a well-defined winding number, adding the non-Hermitian effect transforms it to a topological system with a point gap on the complex plane, characterized by the winding number. The non-Hermitian SSH3 model exhibits two types of localized states and one zero-energy state, but they are associated with the symmetries. The total Zak phase of the non-Hermitian SSH3 model is quantized and indicates where both types of localized states coexist. The SSH4 model is, in many aspects, the original SSH model in disguise. Nevertheless, the non-Hermitian SSH4 model exhibits the non-Hermitian skin effect, causing the skewed profiles of the bulk states. By considering the generalized Brillouin zone or the real-space winding number, the bulk-boundary correspondence of the non-Hermitian SSH4 model is restored. The phenomena presented here may be realized in cold-atom systems or other simulators, and the results will offer more examples of interesting non-Hermitian topological systems.

\begin{acknowledgments}
Y. H was supported by the National Natural Science Foundation of China under Grant No. 11874272.
\end{acknowledgments}

\bibliographystyle{apsrev}

\end{document}